\documentclass[pre,showpacs,onecolumn,preprintnumbers,superscriptaddress,amsmath,amssymb,color,amsfonts]{revtex4-1}
\usepackage{graphicx}
\usepackage{caption}    
\usepackage{subcaption} 

\usepackage{epstopdf}
\bibstyle{apsrev4-1}
\usepackage{color}

\usepackage{mathrsfs}

\usepackage[caption=false]{subfig}

\newsavebox{\astrutbox}
\sbox{\astrutbox}{\rule[-5pt]{0pt}{20pt}}

\begin{document}
\title{ Hydrodynamic interaction in Lattice Fokker Planck  for dilute polymer dynamics }
 \author{Shiwani Singh}
 \email{singh.shiwani@gmail.com}
  \affiliation{School of Computing and Data Sciences, FLAME University, Pune, 412115, India}
  
 \author{ Santosh Ansumali}
 \affiliation{Engineering Mechanics Unit, Jawaharlal Nehru Centre for Advanced
Scientific Research, Jakkur, Bangalore, 560064, Karnataka, India.}

\begin{abstract}
We extend the BGK-type kinetic model for polymer dumbbells and its lattice Boltzmann implementation by incorporating hydrodynamic interaction (HI) effects to accurately recover the slow-variable dynamics governed by the Smoluchowski equation. Unlike Brownian dynamics approaches, which demand the computationally expensive evaluation of the square root of the configuration-dependent diffusion tensor at every time step, this kinetic method is straightforward to implement and avoids such overhead entirely. Consequently, it achieves a significant reduction in computational cost. We validate the model in steady and transient shear flows, demonstrating its ability to capture the distinct signatures of hydrodynamic interactions on bulk rheological properties.
 \end{abstract}

\maketitle
\section{\label{Introduction}Introduction}
 Polymer solutions exhibit complex non-equilibrium behavior driven by the interplay between microscopic configurational dynamics and macroscopic flow fields.  Beyond drag and intramolecular spring forces, each bead in a polymer modeled as a dumbbell experiences additional solvent-mediated interactions that significantly influence the material response of the solution. 
 One such effect is the hydrodynamic interaction (HI) between beads, which arises from the disturbance velocity induced by the movement of other beads \cite{kirkwood1948intrinsic,doi1988theory,bird1987dynamics}.
  Hydrodynamic interactions are known to significantly alter the relaxation spectra of dilute polymer solutions and are crucial for capturing the correct molecular weight dependence of polymer diffusivity and its contribution to shear viscosity. 
 It also influence bulk rheological properties; for instance, they lead to a nonzero second normal stress difference in the FENE model \cite{prakash1999kinetic,ottinger1996stochastic}.
The effects of  HI in canonical flow setups, such as shear flow  are well understood
\cite{ottinger1989gaussian,zylka1989comparison,rudisill1992brownian,prakash1999kinetic,kailasham2023shear,varakhedkar2026linear}. Consequently,  recent  focus is on understanding their role in turbulent and random flows \cite{picardo2025turbulent,ganesh2026hydrodynamic}.
These disturbance velocities are usually modeled using the Oseen-Burgers tensor ${\pmb \Omega}$, with the corresponding configuration-dependent diffusion tensor expressed in terms of Kronecker delta ${\pmb \delta}$ as $({\pmb \delta}-{\pmb \Omega})$  \cite{oseen1927neuere,bird_vol_2}.
 The Oseen tensor captures the long-range hydrodynamic interactions between beads, leading to an anisotropic mobility that profoundly influences the dynamical evolution of the system. 
Incorporating conformation-dependent HI into traditional stochastic solvers, such as Brownian dynamics (BD) simulations, is inherently expensive. This central computational bottleneck arises because the algorithm must evaluate the square root of the configuration-dependent diffusion tensor at every time step \cite{zylka1989comparison,ottinger1996stochastic}.


On the other hand, to avoid the computational burden of computing the square root of the diffusion tensor, it is preferable to work directly with the underlying Smoluchowski equation governing the configuration probability density. In recent years, several methodologies have emerged as efficient solvers for this equation. In these approaches, the fast momentum-relaxation dynamics of polymer molecules—governed by the Fokker–Planck equation—are replaced by a BGK-type relaxation, which preserves the underlying slow dynamics in configuration space \cite{singh2013lattice,singh2011lattice,ammar2010lattice}.
Subsequently, building on this idea,  kinetic-level coupling strategies between solvent and polymer have been developed using the lattice Fokker–Planck method \cite{singh2022two}

This, in turn, motivates the exploration of computational frameworks that retain the efficiency of kinetic-level formulations while extending their applicability to more complex  phenomena, such as hydrodynamic interactions. In this work, we propose an LB-based scheme that accounts for hydrodynamic interactions between  beads through an appropriate modification at the kinetic level. This approach eliminates the need to compute the square root of the diffusion tensor by directly discretizing the underlying Smoluchowski equation.

The paper is organized as follows. In Sec. II, we review the Smoluchowski equation and its equivalent Langevin description for the FENE dumbbell model with hydrodynamic interactions. 
 Furthermore, we discuss the commonly used Oseen–Burgers tensor and its limitations, followed by the introduction of the Rotne–Prager–Yamakawa (RPY) tensor, which overcomes these shortcomings. In Sec. III, we present a modification  of the BGK relaxation mechanism to incorporate hydrodynamic interactions and demonstrate that it reproduces the correct slow dynamics of the configuration probability density. Section IV   outlines the corresponding discrete formulation.    In Sec. V, we compare the present scheme with Brownian dynamics simulations for shear flow, and investigate the impact of hydrodynamic interactions on rheological properties. Finally, Sec. VI concludes the work by demonstrating the computational efficiency of the proposed scheme in terms of CPU time.
\section{ \label{Smolu_hyd}Smoluchowski Equation with  hydrodynamic interactions}
In    the  dumbbell   model,  two beads of equal mass $m_{\rm B}$   are connected by a spring. 
In order to give a  realistic representation of a polymer,
a finitely extensible nonlinear elastic (FENE) spring is preferred over linear infinitely extensible Hookean chain.
 The FENE spring force  ${\bf F}^c $ is given as
 \begin{equation}
 {\pmb F}^c \!=\! \frac{{\rm H} {\pmb Q}}{[1-{Q^2}/{Q_{0}^2}]},
\end{equation}
where ${\pmb Q}$ is the end to end distance between the beads, 
   ${\rm H}$ is the Hookean spring constant and $Q_0$ is the maximum extensibility. 
The equation of motion for the $i^{\rm th}$ bead is
\begin{equation}
 m_{\rm B} \frac{d}{dt} \dot{{\bf R}}_i(t) = {\pmb F}_{i}^c(t) + {\pmb F}_{i}^{\rm D}(t) + {\pmb F}_{i}^{\rm B}(t),
 \label{Langevin}
\end{equation}
where ${\bf R}_i$ denotes the position of the $i^{\rm th}$ bead and ${\pmb F}_{i}^{\rm B}$ represents the Brownian force. The spring forces satisfy the condition ${\pmb F}_1^c = -{\pmb F}_2^c = {\pmb F}^c$.   The total hydrodynamic drag force exerted by the solvent on the $i^{\rm th}$ bead is assumed to be
\begin{equation}
 {\pmb F}_{i}^{\rm D} = -\zeta \,(\dot{{\bf R}}_i - ({\pmb u}_i + \Delta {\pmb u}_i)),
\end{equation}

where $\zeta = 6\pi\eta_{s} a$ is the friction coefficient for a spherical bead of radius $a$, and $\eta_s$ is the solvent viscosity. The total velocity of the solvent at the position of the $i^{\rm th}$ bead is $({\pmb u}_i + \Delta {\pmb u}_i)$, where the imposed velocity is ${\pmb u}_i = {\pmb u}_0 + {\pmb \kappa} \cdot {\bf R}_i$, with ${\pmb u}_0$ as the mean velocity and ${\pmb \kappa}$ as the transpose of the velocity gradient tensor. 
The perturbation $\Delta {\pmb u}_i$ represents the velocity field at ${\bf R}_i$ induced by the hydrodynamic forces exerted by the other beads. Specifically, the perturbation at location $i$ due to a bead at location $j$ ($j \neq i$) is
\begin{equation}
 \Delta {\pmb u}_i = {\pmb \Omega}({\bf R}_i - {\bf R}_j) \cdot (-{\pmb F}_j^{\rm D}),
\end{equation}
where ${\pmb \Omega}$ is the hydrodynamic-interaction tensor.    
  In the inertia-less (low Reynolds number) limit, where $\sum {\bf F} \approx 0$, the drag force balances the spring and Brownian forces (${\pmb F}_i^{\rm D} = - ({\pmb F}_i^c + {\pmb F}_i^{\rm B})$), allowing us to write
\begin{equation}
 \Delta {\pmb u}_i = {\pmb \Omega}({\bf R}_i - {\bf R}_j) \cdot ({\pmb F}_j^c + {\pmb F}_j^{\rm B}).
 \label{pert}
\end{equation}

It is common in the literature on hydrodynamic interactions to use the Oseen–Burgers tensor as a first-order approximation \cite{oseen1927neuere,bird_vol_2}:
\begin{equation}
 {\pmb \Omega}_{\rm Oseen} = \frac{1}{8\pi\eta_s Q} \left({\pmb \delta} + \frac{{\pmb Q}{\pmb Q}}{Q^2}\right).
\end{equation}
The disturbance flow generated by one bead is therefore transmitted through the solvent to the other bead via the hydrodynamic-interaction tensor. Substituting Eq. \eqref{pert} into the force balance Eq. \eqref{Langevin} gives the generalized Langevin equation in the inertialess limit:

 \begin{equation}
   \frac{d{\bf R}_i}{dt} = {\pmb u}_i + \Delta {\pmb u}_i + \frac{1}{\zeta \,}\left({\pmb F}_{i}^c(t) + {\pmb F}_{i}^{\rm B}(t)\right),
\end{equation}
\begin{equation}
\label{finalParticle}
\implies \frac{d{\bf R}_i}{dt} = {\pmb u}_0 + {\pmb \kappa} \cdot {\bf R}_i + {\pmb \Omega}({\bf R}_i - {\bf R}_j) \cdot ({\pmb F}_j^c + {\pmb F}_j^{\rm B}) + \frac{1}{\zeta} \left( {\pmb F}_i^c + {\pmb F}_i^{\rm B} \right). \end{equation}
 The Brownian force, $F^B_i$, which arises from the random collisions of solvent molecules with the beads of the dumbbell, obeys Gaussian statistics, with mean and variance given by:
 \begin{equation}
 \langle F^B_
i (t)\rangle = 0, \langle F^B_
i (t)F^B_
j (t^\prime)\rangle= 2k_B T\zeta \delta_{ij} \delta(t -t^\prime) {\pmb \delta},
 \end{equation}
where $k_B$ is the Boltzmann constant, $T$ is the absolute temperature, ${\pmb \delta}$
 is the Kronecker delta and $\delta(t-t^\prime)$  is the Dirac delta function. In the present study, we focus on a polymer dumbbell and formulate its dynamics in terms of the center-of-mass coordinate, ${\bf R}=({\bf R}_1+{\bf R}_2)/2$, and the connector (end-to-end) vector,
${\bf Q}={\bf R}_2-{\bf R}_1$.
In homogeneous flows, the center-of-mass dynamics decouples from the configurational dynamics. Therefore,  the configurational dynamics is completely described by the evolution of the end-to-end vector ${\bf Q}$, which using Eq.\ref{finalParticle} is\cite{ottinger1996stochastic,zylka1989comparison}: 
 \begin{equation}
 \label{stochastic}
d{\pmb  Q} =\left[
{\pmb \kappa}\cdot{\pmb Q}   - \frac{2H}{\zeta}
[{\pmb \delta}-
 \zeta{\pmb \Omega}({\pmb Q})]\cdot\frac{  \pmb Q  }{1-\frac{Q^2}{Q_0^2}}\right]dt+\sqrt{dt}\,{ \pmb B}\cdot {\pmb \lambda}(t),
\end{equation}
where ${\pmb \lambda}(t)$ is the Gaussian random variables which is
completely characterized by
the first two correlation functions as
\begin{equation}
 \langle{\pmb \lambda}(t)\rangle={\pmb 0},\quad \langle{\pmb \lambda}(t)\,{\pmb \lambda}(t^{\prime})\rangle={\pmb \delta}\,
\delta(t-t^{\prime}),\
\end{equation}
and the configuration-dependent tensor ${\pmb B}$ has to satisfy the following condition:
\begin{equation}
 {\pmb B}\cdot{\pmb B}^{\rm T}=\frac{4 k_{\rm B}T}{\zeta} ({\pmb \delta}-\zeta{\pmb \Omega}),
\label{b_rel}
\end{equation}
with ${\pmb B}^{\rm T}$ standing for  the transpose of ${\pmb B}$.
For a given ${\pmb \Omega}$, i.e., for a given conformation of the system,
${\pmb B}$ is not unique but any ${\pmb  B}$ which satisfying Eq. \eqref{b_rel} can be used
in Eq. \eqref{stochastic}.
One convenient choice for ${\pmb B}$ is obtained from the Cholesky decomposition of  [$\frac{4 k_{\rm B}T}{\zeta} ({\pmb \delta}-\zeta{\pmb \Omega})$].

  Oseen-Burgers tensor, ${\pmb\Omega}_{\rm Oseen}$,
suffers from a drawback that it leads to a nonpositive-definite diffusion tensor for small distances \cite{oseen1927neuere,rotne1969variational,yamakawa1970transport}.
The Brownian dynamics simulations for the Oseen-Burgers tensor are  possible  
only for positive-definite diffusion tensors. Therefore,
hydrodynamic-interaction tensor is chosen such that 
 the diffusion tensor $({\pmb \delta}-{\pmb \Omega})$ is always
positive-definite for all configurations. 
In order to remove this problem, the Rotne-Prager-Yamakawa (RPY) tensor 
was introduced which
modifies the small separation disturbance in such a manner 
that the diffusion tensor is always positive-definite \cite{rotne1969variational,yamakawa1970transport}. 
 The Rotne-Prager-Yamakawa tensor has    two parts depending on the extension of the
dumbbell as:
 \begin{eqnarray}
 {\pmb \Omega}_{\rm RPY}({\pmb Q})=\frac{1}{8\pi\eta_s Q}
\begin{cases} \left(1+\frac{2a^2}{3Q^2}\right){\pmb \delta}+\left(1-\frac{2a^2}{Q^2}\right)\frac{{\pmb Q}{\pmb Q}}{Q^2} &\textrm{ for }Q\geq 2a\\
  \frac{Q}{2a}\left(\frac{8}{3}-\frac{3Q}{4a}\right){\pmb \delta}+\frac{Q}{4a}\frac{{\pmb Q}{\pmb Q}}{Q^2} &\textrm{ for }Q< 2a.
\end{cases}
\end{eqnarray}
Throughout  our study onward, we will be using ${\pmb \Omega}^{\rm RPY}$ to model hydrodynamic interaction.
Having specified the form of the hydrodynamic interaction tensor through ${\pmb \Omega}^{\rm RPY}$, we now shift our focus from the stochastic trajectory description of the connector vector (${\pmb Q}$) to an equivalent probabilistic description in terms of the distribution function ($\psi({\pmb Q},t)$ ) which gives the probability of finding ${\pmb Q}$ at time ($t$).
The Smoluchowski equation governing the dynamics of the configuration distribution function $\psi({\pmb Q},t)$ 
in the presence of the hydrodynamic interaction
is given as \cite{bird_vol_2}:
\begin{equation}
\frac{\partial \psi}{\partial t} + \frac{\partial}{\partial Q_{\alpha}} \left\{{\pmb \kappa}\cdot{\pmb Q}\psi-
\frac{2H}{\zeta}
\left({\pmb \delta} -     \zeta {\pmb \Omega}^{\rm RPY}\right)\cdot \frac{  \pmb Q  }{1-\frac{Q^2}{Q_0^2}} \psi\right\} 
  \!=\! \frac{2k_{\rm B} T_0}{\zeta} \frac{\partial  }{\partial {\pmb Q}}\cdot\left({\pmb \delta} -  
\zeta{\pmb  \Omega}\right)\cdot\frac{\partial \psi}{\partial {\pmb Q}},
\label{Smoluchowski}
\end{equation}
Equation \eqref{Smoluchowski} can now be written in non-dimensional form by scaling lengths with the equilibrium length $l_0=\sqrt{k_bT/{H}}$ and time with  the one associated with the restoring action of spring force, $\tau_{\rm R}=\zeta/(4H)$. The final non-dimesional Smoluchowski equation has the following form:
\begin{equation}
\frac{\partial \psi}{\partial t} + \underbrace{\frac{\partial}{\partial Q_{\alpha}} \left\{{\rm Wi}\,{\pmb \kappa}\cdot{\pmb Q}\psi-
\frac{1}{2}
\left({\pmb \delta} -     {\pmb \Omega}^{\rm RPY}\right)\cdot \frac{  \pmb Q  }{1-\frac{Q^2}{b}} \psi\right\} }_{\it drift}
  \!=\! \underbrace{\frac{1}{2} \frac{\partial  }{\partial {\pmb Q}}\cdot\left({\pmb \delta} -  
{\pmb  \Omega^{\rm RPY}}\right)\cdot\frac{\partial \psi}{\partial {\pmb Q}}}_{\it diffusin},
\label{Smoluchowski_nondim}
\end{equation}
where Weissenberg number  is defined as ${\rm Wi}=\tau_{\rm R}\kappa$ with $\kappa$ now being the the magnitude of the characteristic shear rate. The parameter $b=Q_0/l_0$ is called the maximum extensibility parameter.    The Rotne-Prager-
Yamakawa tensor has the following non-dimensinal form:  
\begin{eqnarray}
 {\pmb \Omega}_{\rm RPY}({\pmb Q})=\frac{3}{4}h^{\star}\frac{1}{Q}
\begin{cases} \left(1+\frac{2a^2}{3Q^2}\right){\pmb \delta}+\left(1-\frac{2a^2}{Q^2}\right)\frac{{\pmb Q}{\pmb Q}}{Q^2} &\textrm{ for }Q\geq 2a\\
  \frac{Q}{2a}\left(\frac{8}{3}-\frac{3Q}{4a}\right){\pmb \delta}+\frac{Q}{4a}\frac{{\pmb Q}{\pmb Q}}{Q^2} &\textrm{ for }Q< 2a.
\end{cases}
\label{non-dimRPY}
\end{eqnarray}
where $h^{\star}=3a/4 $ with non-dimensional bead radius, $a$. 
The parameter $h^{\star}$  is
roughly the ratio of bead radius  and equilibrium extension and quantifies the strength of solvent-mediated hydrodynamic coupling between beads.  The ${\it drift}$ and ${\it diffusion}$ terms of Eq. \eqref{Smoluchowski_nondim}  both contains the configuration dependent hydro-dynamic interaction tensor.

\section{Kinetic Formulation \label{formulation}}
In Refs. \cite{singh2013lattice,singh2011lattice,ammar2010lattice,singh2022two}, it has been shown that the  diffusive dynamics governing the momentum relaxation
of a polymer molecule in phase space can   be replaced with a relaxation mechanism of BGK type without 
altering the actual slow dynamics in configuration space.
  This approach utilizes the fact that momentum is conserved for the system and not for individual component.   
Hence, for the rheological flows, this fact is used to treat the velocity field of the given set-up, such as shear flow, extensional flows etc,  as an imposed  velocity field via  BGK relaxation mechanism. There are two main time-scales in this problem.  The first one is the  time scale associated with the
momentum relaxation of the polymer molecule due to random kicks (collisions)  from the surrounding fluid molecules which is $\tau_{\rm inertial} = m_{\rm B} /\zeta$ and the and the second one  is    associated with the restoring action of spring force used to model polymer molecule,
$ \tau_{\rm R}= \zeta/(4H)$.
  Due to typical gap in these two time-scales, one of the obvious small parameters of this system  is the ratio of  these two time scales $\epsilon=\tau_{\rm inertia}/\tau_{\rm R}$.

Before describing model for hydrodynamic interaction, we  briefly overview the kinetic model for dilute polymer solutions without the inclusion of HI, as presented in Refs. \cite{singh2013lattice,singh2011lattice}. We use $\tau_{\rm R}$ and  $\sqrt{(2k_{\rm B}T)/m_{\rm B}}$ for  
time and velocity scale, respectively, to obtain the non-dimensional form of the BGK-Boltzmann equation in a phase space governed by polymer configurations as
  \begin{align}
\begin{split}
  \frac{\partial}{\partial t}  f ({\pmb Q},\dot{\pmb Q},t)  + { \dot{  Q}_{\alpha}}    \frac{\partial  }{\partial{   { Q}_{\alpha} }}f ({\pmb Q},\dot{\pmb Q},t)
 =\frac{1}{\epsilon}[
f^{ \rm eq}( {\psi}({\pmb Q},t),\bar{\pmb v},T)
-f ({\pmb Q},\dot{\pmb Q},t) ].
\label{single_relax}
\end{split}
\end{align}
The kinetic equation (Eq. \eqref{single_relax} describes the dynamics of an ensemble of polymer molecules at the microscopic level. Each polymer molecule is characterized by its end-to-end vector (${\pmb Q}$), which represents its instantaneous configuration, and by the  corresponding configuration velocity ($\dot{\pmb Q}$), which describes how this ${\pmb Q}$ changes with time. The distribution function $f({\pmb Q},\dot{\pmb Q},t)$ therefore gives the probability density of finding polymer molecules with a particular configuration and configuration velocity at a given time. The BGK term, on the right-hand side of Eq. \eqref{single_relax}, models the relaxation of the polymer distribution function $f$ toward the local equilibrium distribution $f^{\rm eq}$ at a rate controlled by $\epsilon$. Finally, the configuration probability density $\psi({\pmb Q},t)=\int f({\pmb Q},\dot{\pmb Q},t)d\dot{\pmb Q}$ describes how the polymer molecules are distributed over different configurations.
The relevant equilibrium distribution is given as
\begin{equation}
f^{eq} =\psi \left(\frac{1}{2 \pi  }\right)^{\frac{3}{2}}\exp\left[-\frac{ (\dot{ {Q}}_{\alpha}- { \bar{ v}_{\alpha}})^2}{2}\right],
\label{f^eq}
\end{equation}
with  the imposed velocity $\bar{ v}_\alpha$, with the following form:
\begin{align}
\begin{split}
  \bar{ v}_\alpha({\pmb Q}, t)
  &=  \left({\rm Wi}\,{\kappa}_{\alpha\beta}{Q_\beta} -\frac{1}{2} 
{\ F}^{c}_\alpha\right)
\label{imp_vel}
\end{split}
\end{align}
where  ${F}^c_\alpha$ is the non-dimensional spring force given as ${ Q_\alpha}/(1-\frac{Q^2}{b}) $.
%
  The first and second order moments of the phase space density are defined as
 \begin{equation}
     \psi u_\alpha=\int \dot{Q}_\alpha f({\pmb Q, \dot{\pmb Q}, t}) d\dot{\pmb Q}, \qquad P_{\alpha\beta}=\int \dot{Q}_\alpha \dot{Q}_\beta f({\pmb Q, \dot{\pmb Q}, t}) d\dot{\pmb Q}.
 \end{equation}

To understand slow dynamics, we write the moment chain of Eq\eqref{single_relax} as  
\begin{align}
\begin{split}
  \frac{\partial  \psi}{\partial t}+
  \frac{\partial \psi u_\alpha}{\partial Q_\alpha}
  &=  0,\\
    \frac{\partial\psi u_\alpha  }{\partial t}+
  \frac{\partial P_{\alpha\beta}}{\partial Q_\beta}
  &=  \frac{1}{\epsilon}(\psi \bar{v}_\alpha-\psi u_\alpha),\\
  \frac{\partial P_{\alpha\beta} }{\partial t}  + 
  \frac{\partial M_{\alpha\beta\gamma}}{\partial Q_\gamma}
  &= \frac{1}{\epsilon}(\bar{P}^{eq}_{\alpha\beta}-P_{\alpha\beta}), \\
\end{split}
\end{align}
where the third order moment $M_{\alpha\beta\gamma}=\int \dot{Q}_\alpha \dot{Q}_\beta \dot{Q}_\gamma  f({\pmb Q, \dot{\pmb Q}, t}) d\dot{\pmb Q}$ and $\bar{P}^{eq}_{\alpha\beta}=\psi \delta_{\alpha\beta}+\psi\bar{v}_\alpha\bar{v}_\beta$.
The first moment dynamics here reveal that 
the first moment  is relaxing toward  an imposed value set by the external velocity field. Therefore, momentum, as obtained by the first order moment, can be expanded as $\psi{ u}_\alpha=\psi \bar{ v}_\alpha+\epsilon\psi{ u}_\alpha^{(1)}+...$. Furthermore, on 
 doing a Chapman-Enskog expansion around equilibrium given by Eq. \eqref{single_relax}, the dynamics at O(1) results with a diffusion term in the   evolution equation configuration probability density $\psi$ as  (see Ref.\cite{singh2013lattice}):
 \begin{equation}
 \frac{\partial \psi}{\partial t}+\frac{\partial( \psi \bar{v}_\alpha) }{\partial Q_\alpha}=\epsilon\frac{\partial^2 \psi}{\partial Q^2}.
 \end{equation}
 Finally, using the relevant length-scale in the configuration space, which is the equilibrium extension $l_0$ instead of the one associated with the kinetic scale, the Smoluchowski equation takes the following form:
 \begin{equation}
   \frac{\partial \psi}{\partial t}+\frac{\partial}{\partial Q_\alpha}\left[ \psi\left({\rm Wi}\,{\kappa}_{\alpha\beta}{ Q_\beta} -\frac{1}{2} 
{ F_\alpha}^{c}\right)\right]=\frac{1}{2}\frac{\partial^2 \psi}{\partial Q^2}.  
 \end{equation}
 The kinetic equation \eqref{single_relax} was solved using a lattice Boltzmann method formulation Ref.\cite{singh2013lattice}.

Our objective    in this work is to  build a similar kinetic  formulation to include the effects of hydro-dynamic interaction. Apriori, we know that  the slow dynamics for  configuration density is as given by the HI-adjusted Smoluchowki Eq. \eqref{Smoluchowski_nondim} with ${\it drift}$ and ${\it diffusion}$ term containing the effect of hydrodynamic-interation tensor. Motivated by the approach in Ref. \cite{singh2013lattice,singh2011lattice}, we seek a modification at the kinetic level that yields a hydrodynamic-interaction-adjusted Smoluchowski equation (Eq. \eqref{Smoluchowski_nondim}) at $O(1)$ dynamics.
 Similar to Ref \cite{singh2013lattice},  we  begin with a single-relaxation-time BGK equation  but with a modified equilibrium distribution function given by   a quasi-equilibrium  Grad like distribution \cite{grad1949kinetic}. Hence, the governing equation at the kinetic level is:
\begin{align}
\begin{split}
  \frac{\partial}{\partial t}  f ({\pmb Q},\dot{\pmb Q},t)  + { \dot{  Q}_{\alpha}}    \frac{\partial  }{\partial{   { Q}_{\alpha} }}f ({\pmb Q},\dot{\pmb Q},t)
 =\frac{1}{\epsilon}[
f^{ \rm Grad}
-f ({\pmb Q},\dot{\pmb Q},t) ].
\label{single_relax_grad}
\end{split}
\end{align}
The classical Grad’s distribution function \cite{grad1949kinetic} was introduced to approximate the solution of the Boltzmann equation by expanding the distribution function in terms of a finite set of moments using Hermite polynomials, thereby incorporating higher-order moments—such as stress and heat flux—to capture non-equilibrium effects beyond the conserved fields.
For the purpose of present case, the Hermite expansion is truncated at the second order. Therefore, the Grad’s distributions are also parameterized by the values of stress-tensor (see Appendix B). 
The motivation for this choice is the anticipation that it will give rise to a tensorial {\it diffusion} term in the O(1) dynamics upon performing the Chapman–Enskog expansion.
Accordingly, the relevant non-dimensional form of Grad’s distribution for the polymer dumbbell case is given by:
\[
f^{\mathrm{Grad}} =
\left(\frac{1}{2\pi}\right)^{3/2}
\exp\!\left[
-\frac{   \dot{Q}_\alpha^2}{2}
\right]
\left[
\psi+\psi v^{\rm HI}_\alpha \dot{Q}_\alpha
+ \frac{1}{2}\,(\Theta_{\alpha\beta}-\psi\delta_{\alpha\beta})
 (\dot{Q}_\alpha \dot{Q}_\beta-\delta_{\alpha\beta})
\right].
\]
Choosing the following form of the imposed velocity effectively incorporates the HI tensor into the {\it drift} term:
\begin{align}
\begin{split}
  { v}^{\rm HI}_\alpha({\pmb Q}, t)
  &=  \left({\rm Wi}\,{\kappa}_{\alpha\beta}{ Q_\beta} -\frac{1}{2}\{{ \delta}_{\alpha\beta}-{\ \Omega}^{\rm RPY}_{\alpha\beta}({\pmb Q})\}
{ F}^{c}_\beta\right),  
\label{imp_vel_hyd}
\end{split}
\end{align}
Furthermore, on choosing ${ \Theta_{\alpha\beta}}=\psi({\delta_{\alpha\beta}}-{ \Omega_{\alpha\beta}}^{\rm RPY})$,
the moment equation for the kinetic equation \eqref{single_relax_grad}:
\begin{align}
\begin{split}
  \frac{\partial  \psi}{\partial t}+
  \frac{\partial \psi u_\alpha}{\partial Q_\alpha}
  &=  0,\\
    \frac{\partial\psi u_\alpha  }{\partial t}+
  \frac{\partial P_{\alpha\beta}}{\partial Q_\beta}
  &=  \frac{1}{\epsilon}(\psi {v}^{\rm HI}_\alpha-\psi u_\alpha),\\
  \frac{\partial P_{\alpha\beta} }{\partial t}  + 
  \frac{\partial M_{\alpha\beta\gamma}}{\partial Q_\gamma}
  &= \frac{1}{\epsilon}(\psi\delta_{\alpha\beta}-\psi\Omega^{\rm RPY}_{\alpha\beta}-P_{\alpha\beta}), \\
\end{split}
\label{moment_eq}
\end{align}
Subsequently, performing a Chapman–Enskog expansion (see Appendix A) yields the desired slow dynamics at O(1) as follows:
\begin{equation}
 \frac{\partial \psi}{\partial t}+\frac{\partial( \psi {v}^{\rm HI}_\alpha) }{\partial Q_\alpha}=\epsilon\frac{\partial }{\partial Q_\alpha}\left(\delta_{\alpha\beta}-\Omega^{\rm RPY}_{\alpha\beta}\right)
 \frac{\partial \psi}{\partial Q_\beta}.
 \end{equation}
Finally, on choosing the equilibrium extension as the characteristic length scale, the non-dimensional Smoluchowski equation takes the form of the desired HI-adjusted evolution equation for the configuration density (Eq. \eqref{Smoluchowski_nondim}):
\begin{equation}
\frac{\partial  {\psi}}{\partial t} + \frac{\partial}{\partial { Q}_\alpha } \cdot\left\{  
\psi\, {\rm Wi}\,\,{\kappa}_{\alpha\beta}{ Q}_\beta     - \frac{1}{2}
\left({ ( \delta}_{\alpha\beta}-
\zeta { \Omega}_{\alpha\beta}^{\rm RPY}({\pmb Q}))\frac{  Q_\beta  }{1-\frac{Q^2}{b}}
   \right){\psi}\right\} 
  \!=\! \frac{1 }{2}\frac{\partial  }{\partial {Q_\beta}}\cdot({ \delta_{\alpha\beta}}-
\zeta { \Omega_{\alpha\beta}}^{\rm RPY}({\pmb Q}))\frac{\partial  {\psi}}{\partial{Q_\beta}}.
\label{smoluchowski_nondim}
\end{equation}

\section{Discrete formulation via lattice Boltzmann method\label{discrete_form}}
 \begin{figure}[h]
\begin{center}
 \includegraphics[scale=0.4]{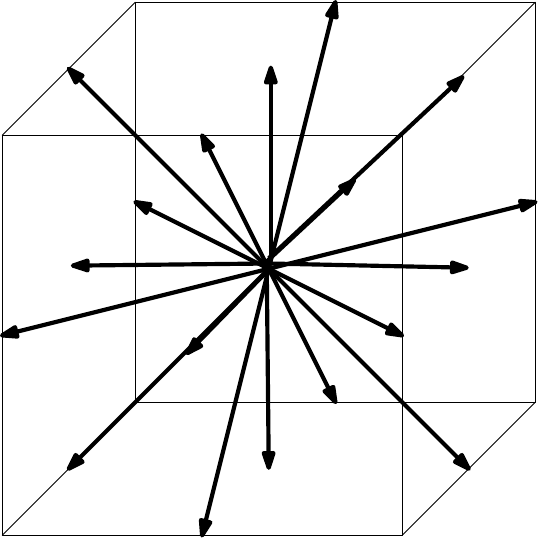}
\caption{ D3Q19 velocity-space model}
\label{vel_space}
\end{center}
\end{figure}
In this section, we will build a discrete solver  based on the lattice Boltzmann method (LBM) to solve the Smoluchowski equation (Eq. \eqref{smoluchowski_nondim}). We emphasize that the use of LBM is not a requirement for solving the Smoluchowski equation, and other numerical schemes can be adopted for this purpose \cite{lozinski2003fast,chauviere2004simulation}. The present choice of LBM is motivated primarily by its natural compatibility with the lattice Boltzmann framework used for the underlying flow field. This allows the the polymer dynamics governed by the Smoluchowski equation to be incorporated into the existing flow solver in a straightforward manner, facilitating an efficient and consistent treatment of the coupled fluid–polymer dynamics.
In  LB  formulation,  one  works  with a set of discrete populations $f=\{f_i\}$ corresponding to the  discrete velocities  defined
 as ${\bf c}_i$ $(i=1,\cdots,N)$ to represent the system \cite{succi_book}. These vectors form the links of a lattice that should satisfy appropriate isotropy condition.
 In the present case, the spatial coordinate $\pmb{x}$ corresponds to the end-to-end vector $\pmb{Q}$, while the velocity coordinate $\pmb{c}$ represents its rate of change $\dot{\pmb{Q}}$. Accordingly, in this context, the discrete velocities are given by $\dot{\pmb{Q}}_i$.
In this study, we employ the D3Q19 velocity-space model, which comprises a discrete velocity set in phase space consisting of simple cubic (SC), face-centered cubic (FCC), and a zero-velocity component as shown in the Fig. \ref{vel_space}.
The weights corresponding  to the structures of  D3Q19  velocity-space model  are 
\begin{equation}
   w_{0}=\frac{1}{3},\quad w_{\rm SC}=\frac{1}{18},\quad w_{\rm FCC}=\frac{1}{36},\quad
 \end{equation}
%
For the  set of discrete populations,   the evolution equation will be written as,
\begin{equation}
\label{two_relax_discrete}
 \frac{\partial f_{i}}{\partial t}+{\dot{Q}}_{\alpha}\frac{\partial    f_{i}}{\partial Q_{\alpha}}=
\frac{1}{\epsilon}[f_{i}^{\rm Grad}(\psi, {\pmb v}^{\rm HI}, {\pmb \Theta})-f_{i} ],
\end{equation}
where the configuration probability density ${\psi}$, the only slow and conserved moment is defined as $\psi = \sum_i f_i$.
Here, $\pmb{v}^{\rm HI}$ denotes an imposed velocity field (see Eq.~\eqref{imp_vel_hyd}), and $\pmb{\Theta}$ is a tensor incorporating hydrodynamic interaction (HI) effects, with the form $\pmb{\Theta} = \psi \left( \pmb{\delta} - \pmb{\Omega}^{\rm RPY} \right)$.
Finally, the discrete equilibrium function is given as (for details see Appendix B)\cite{shan1998discretization,chikatamarla2006grad}:
\begin{equation}
    f_i^{\rm Grad}=\omega_i\left[\psi+\frac{\psi v^{\rm HI}_\gamma \dot{Q}_{i\gamma}}{c_s^2}+\frac{1}{2 c_s^4}(\Theta_{\gamma\theta}-\psi c_s^2\delta_{\gamma\theta})(\dot{Q}_{i\gamma}\dot{Q}_{i\theta}-c_s^2\delta_{\alpha\theta}) \right]
    \label{final_grad_discrete_poly}
    \end{equation}
where $\dot{ Q}= \sqrt{3}c_{\rm s}$ with $c_s= 1/\sqrt{3}$.
After doing time and space discretizations, we
implement the discrete formulation in the following manner for obtaining polymer dynamics in ${\pmb Q}$ space:
\begin{equation}
g_i( {\pmb Q} +\dot{\pmb Q}\Delta t,t+\Delta t)=g_i(  {\pmb Q},t)  +2\beta \left[ g_i^{\rm eq}( \rho, \bar{\pmb  v})  -g_i(  {\pmb Q},t)\right]   
\label{dis_eq}
\end{equation}
where $\beta=\Delta t/(2\epsilon+\Delta t)$ and 
 the auxiliary population $g_i$ is defined as,
\begin{equation}
\label{aux}
 g_i =f_i-\frac{\Delta t}{2\epsilon}\left(f_i-f_i^{\rm eq} \right), 
\end{equation}
In order to avoid the singularity of FENE potential at $Q=\sqrt{b}$, we restrict our simulation domain within a spherical wall of radius $r < \sqrt{b}$. 
At the spherical boundary, a bounce-back condition is imposed such that the outgoing populations are reflected back into the computational domain in the following manner:
\begin{equation}
f({\pmb Q}_w,\dot{\pmb Q}_i,t+\Delta t)=f({\pmb Q}_w,-\dot{\pmb Q}_i,t),
\qquad \qquad {(\dot{\pmb Q}_i\cdot{\pmb n}>0)}\end{equation}
with ${\pmb n}$ being the wall normal.
This prevents the probability distribution function from leaving the domain as wells as ensures mass conservation. The physical basis for using the bounce-back boundary condition is that the configurational probability density \(\psi\) approaches zero near the boundary; hence, the specific form of the boundary condition is not critical, provided that it conserves mass.
The initial value of $\psi$ is given as the equilibrium distribution in the absence of flow \cite{bird_vol_2} which is, 
\begin{equation}
 \psi({\bf Q},0)=
\begin{cases} N_{\rm eq}(1-Q^2/b)^{b/2} & \text{for  $|Q| \leq r$}
\\
0 &\text{ elsewhere.}
\end{cases}
\end{equation}
The solution of equation \eqref{dis_eq} yields the configuration probability density $\psi$ at a given time. Knowledge of $\psi$ is sufficient to evaluate material functions and characterize the influence of hydrodynamic interaction on non-equilibrium behavior of the FENE dumbbell which is discussed
in Sec. \ref{Results}.

\section{\label{Results} Results}
In this section,
 we study the
bulk rheological properties in simple shear flow  to examine  the effect of hydrodynamic interaction on material
function. 
In shear flow  the transpose of velocity gradient is given as:
\begin{equation}
{\pmb \kappa}=\dot{\gamma}\!\begin{bmatrix}
  0 & 1 & 0 \\
  0 & 0 & 0 \\
  0 & 0 & 0
 \end{bmatrix},
\label{trans_shear}
\end{equation}
where $\dot{\gamma} $ is  the shear rate.
The stress tensor for the polymer solution is the sum of a
contribution from the Newtonian solvent and from the polymer which is given as,  ${\pmb \sigma}={\pmb \sigma}_{\rm S}+{\pmb\sigma}_{\rm P}$. The  Newtonian solvent stress,  ${\pmb \sigma}_{\rm S}$, is given as $ \eta_s \dot{\gamma} $ where $\eta_s$ is  the solvent viscosity. The  
Kramers form of the polymer contribution polymeric contribution, ${\pmb \sigma}_{\rm P}$,  to the stress tensor   is given as:
\begin{equation}
{\pmb \sigma}^{\rm P}=n\left\langle {\bf Q}\hat{\pmb
F}^c\right\rangle-n\left\langle {\pmb Q}\hat{\pmb
F}^c\right\rangle_{eq}. \label{exp:polymerstress}
\end{equation}
In simple shear flow set up, one is interested in the following three rheological material functions-viscosity, first normal stress coefficient 
 and the second  normal stress which are respectively defined as
 \begin{align}
\label{mat_fun_sh}
\begin{split}
\frac{\eta(\dot{\gamma})-\eta_{s}}{nk_{\rm B}T\tau_{R}}=\frac{1}{\rm Wi}\left(\frac{\sigma_{xy}}{nk_{\rm B}T}\right),  \quad
\frac{\psi_{1}}{nk_{\rm B}T\tau_{R}^{2}}=\frac{1}{\rm Wi^{2}}\left(\frac{\sigma_{xx}-\sigma_{yy}}{nk_{\rm B}T}\right) \quad \textrm{and}\quad
\frac{\psi_{2}}{nk_{\rm B}T\tau_{R}^{2}}=\frac{1}{\rm Wi^{2}}\left(\frac{\sigma_{yy}-\sigma_{zz}}{nk_{\rm B}T}\right).
\end{split}
\end{align}

%


\begin{figure}
\begin{center}
\includegraphics[scale=0.5]{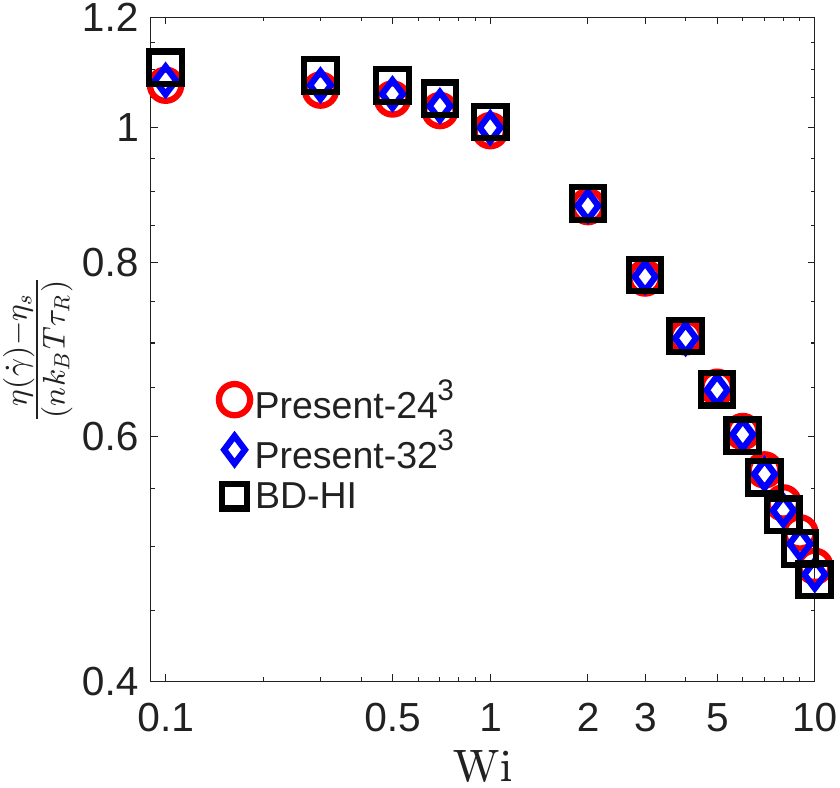}
\includegraphics[scale=0.5]{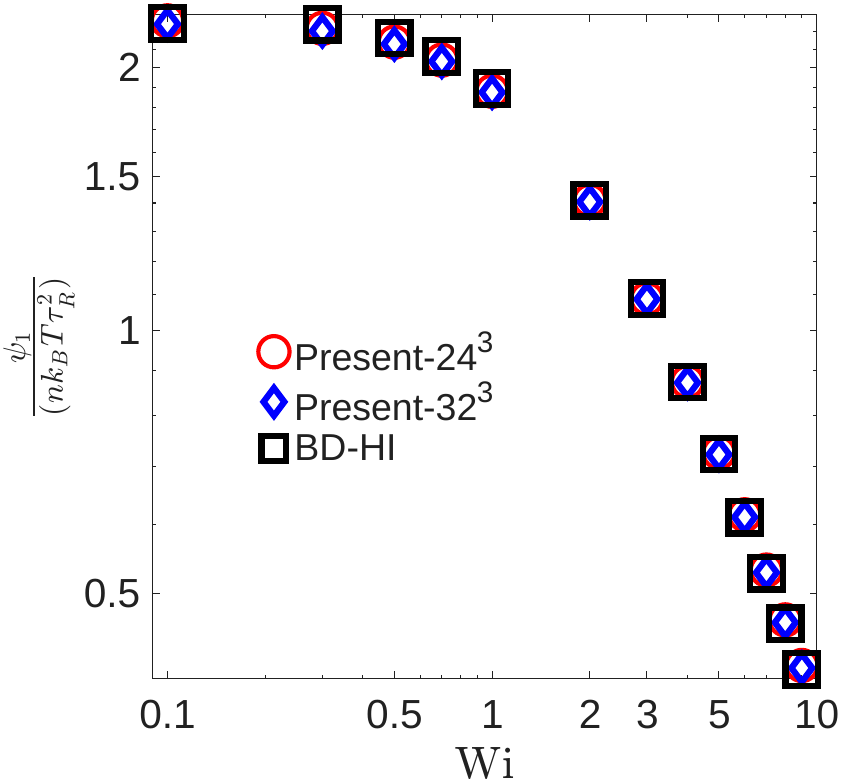}
\caption{Steady state value of the shear viscosity
and the first normal stress coefficient with different Wi for an
extensibility parameter $b = 50$ and $h^{\star}=0.15$. }
\label{Steady state}
\end{center}
\end{figure}
In Fig. \eqref{Steady state}, the shear viscosity and the first normal stress
coefficient for steady simple shear flow are plotted with
respect to Wi for an extensibility parameter $b = 50$ and   hydrodynamic interaction parameter   $h^{\star}=0.15$.
As shown in the figure, the steady-state material function agrees well with the Brownian dynamics (BD) results, which were obtained using a time step of $dt=0.001$ and averaged over $10^7$ trajectories.
\begin{figure}
\begin{center}
\includegraphics[scale=0.5]{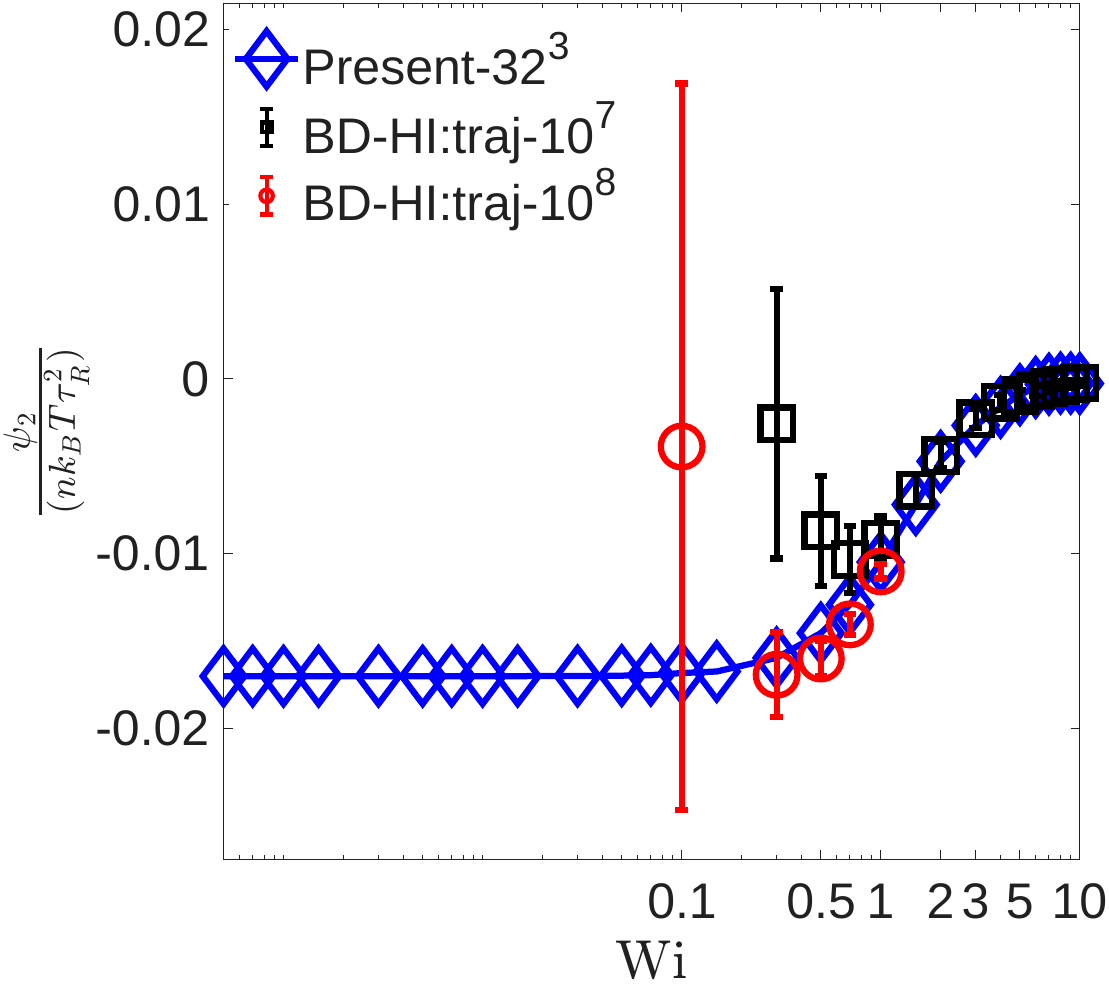}
\caption{  Steady state value of the second normal stress coefficient with different Wi for an
extensibility parameter $b = 50$ and $h^{\star}=0.15$.
}
\label{Steady_state_psi2}
\end{center}
\end{figure}

The steady state value of second normal stress coefficient vanishes in the absence of hydrodynamic interaction for FENE dumbbell. However, in the presence of hydrodynamic interaction,
the steady state value of second normal stress coefficient starts from negative value for small {\rm Wi}
and reaches a saturated value of zero for high Wi number (see Fig. \ref{Steady_state_psi2}). At low Wi when polymers are mostly coiled, the hydrodynamic interactions between the beads leads to an anisotropic interaction with the surrounding fluid. This alters the transverse correlation between the stress resulting in a negative value for the $\psi_2$. 
 
As can be seen from Fig.~\ref{Steady_state_psi2}, the present scheme is capable of accurately predicting this behavior even as ${\rm Wi}\to 0$. However,  when $10^7$ trajectories are used for BD simulations, the statistical errors become significant  around ${\rm Wi}=1$, leading to an inaccurate prediction of the second normal stress difference. Therefore, for ${\rm Wi}\leq 1$, the BD simulations were averaged over $10^8$ trajectories. With this increased number of trajectories, the statistical errors remain reasonably small up to approximately ${\rm Wi}=0.3$. Beyond this (${\rm Wi}=0.3$), however, the error bars indicate that the statistical uncertainty becomes relatively large. In contrast, the present scheme which is  free from such sampling-related statistical fluctuations continues to provide consistent predictions across the entire range of ${\rm Wi}$ considered.



\begin{figure}
\begin{center}
\includegraphics[scale=0.5]{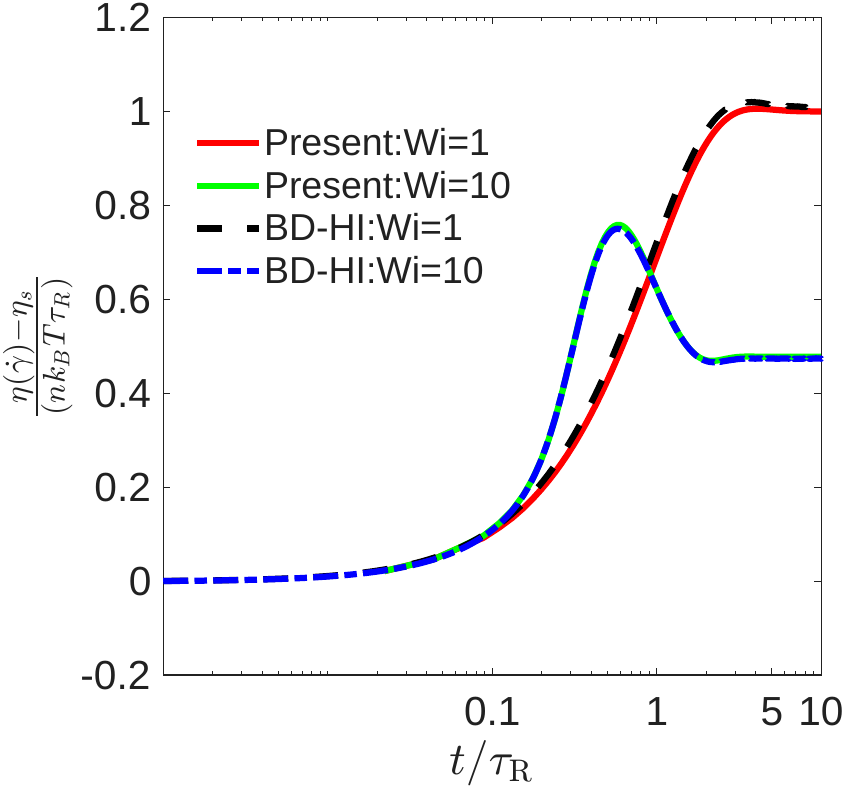}
\includegraphics[scale=0.5]{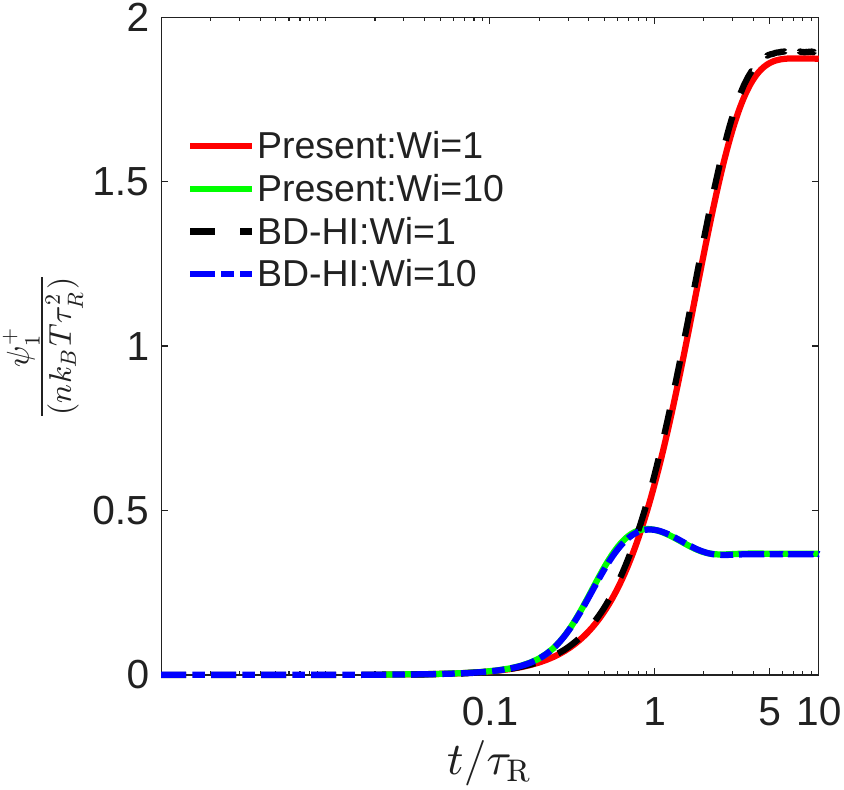}
\caption{  Time evolution of the shear viscosity and the first normal stress coefficient for Wi = 1.0 and 10
($b=50, h^{\star}=0.15$).}
\label{Time dymanics}
\end{center}
\end{figure}
Figure \ref{Time dymanics} presents the time-dependent shear viscosity  and the first normal stress coefficient  for $h^\star=0.15$, at ${\rm Wi=1}$ and 10, and for an extensibility parameter $b=50$, along with a comparison to results obtained from Brownian dynamics simulations.
\begin{figure}
\begin{center}
\includegraphics[scale=0.5]{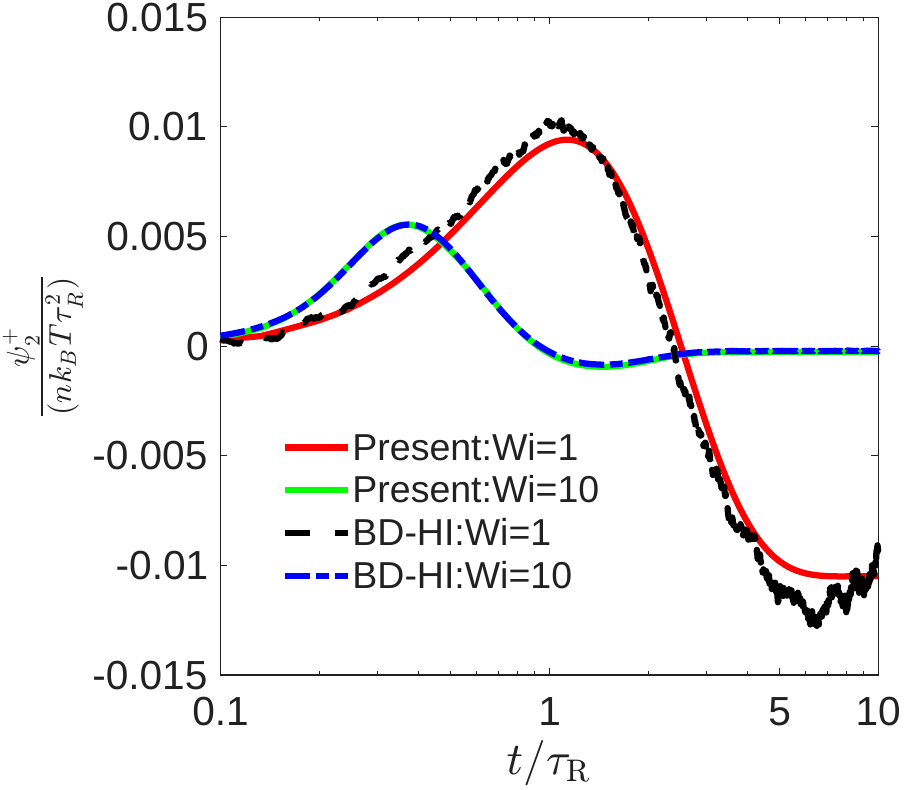}
\caption{Time evolution of the second normal stress coefficient for Wi = 1.0 and 10
($b=50, h^{\star}=0.15$).}
\label{psi_2}
\end{center}
\end{figure}
Further, we present the time dynamics of the second normal stress coefficient $\psi_2^+$ in Fig. \ref{psi_2}. In Fig. \ref{psi_2}, the value  of $\psi_2^+$ for both Wi=1 and Wi=10 returns to a steady state value after the initial overshoot. The steady state value for Wi=10 is close to zero. However, the steady state value for Wi=1 attains a negative value, which is also in accordance with the previous argument used for Fig. \ref{Steady_state_psi2}. 
\begin{figure}
\begin{center}
\begin{subfigure}[b]{0.45\textwidth}
         \centering
\includegraphics[scale=0.5]{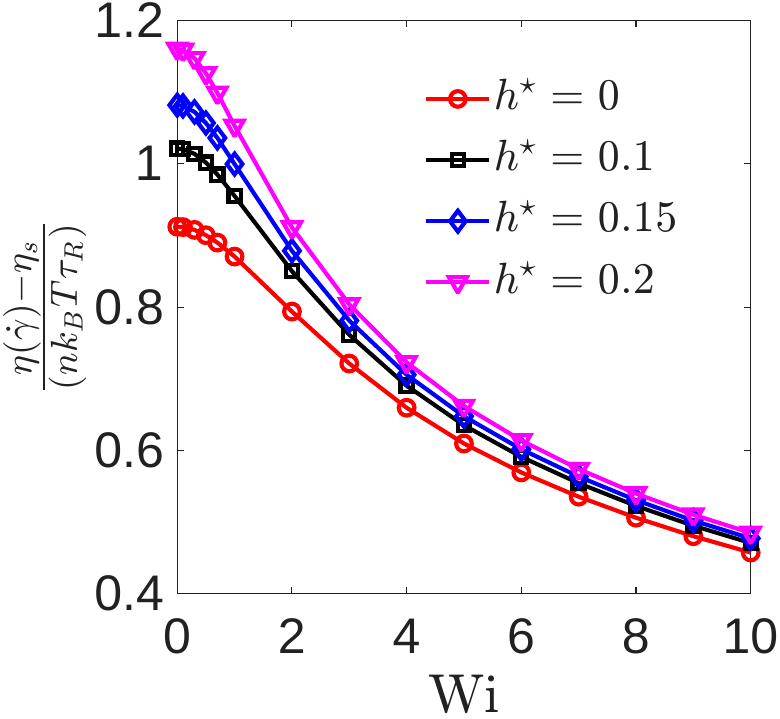}
 \caption{Shear Viscosity}
         \label{sh_vis_diffh}
\end{subfigure}
\begin{subfigure}[b]{0.45\textwidth}
         \centering
\includegraphics[scale=0.5]{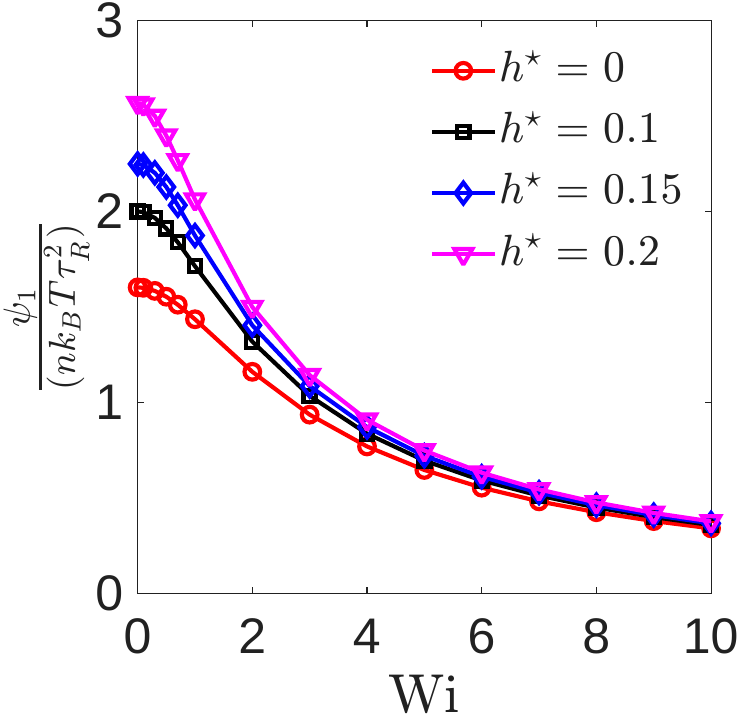}
\caption{First Normal
Stress Coefficient}
         \label{first_nor_diffh}
\end{subfigure}
\caption{Steady state value of the shear viscosity
and the first normal stress coefficient   with respect to different  $h^{\star}$
as function of  Wi for
extensibility parameter $b = 50$. These calculations have been performed with  32 grid points in each direction.}
\label{different_h_star}
\end{center}
\end{figure}


In order to show the effect of hydrodynamic interaction on the shear thinning behavior, we plotted
the steady state value of the shear viscosity and the first normal stress coefficient as a function
of Wi for different $h^{\star}$ in  Fig. \ref{different_h_star}. 
It can be seen from this figure that
the material functions associated with small Wi number are pushed towards higher value as
the strength of hydrodynamic interaction increase.  
This behavior is consistent with the known effect of hydrodynamic interactions, which induce correlated motion between the beads, particularly at lower {\rm Wi}. This enhanced correlation increases the polymer contribution to the stress and slows down stress relaxation, thereby leading to an increase in the zero-shear-rate viscosity of dilute polymer solutions at lower {\rm Wi}. This increase in viscosity towards lower {\rm Wi} with hydrodynamic interactions is also observed in Fig. \ref{sh_vis_diffh} \cite{ottinger1996stochastic,zimm1956dynamics,kailasham2018rheological}.
It is also evident from the figure that the
shear-thinning behavior becomes stronger with increasing $h^{\star}$.  It is know that the finite extensibility becomes particularly significant
to the shear thinning properties of dilute polymer solutions at high shear rate. This fact is evident from Fig. \ref{different_h_star}, where, with increasing ${\rm Wi}$, all the curves tend to converge since the same finite extensibility parameter, $(b=50)$, is used for all the cases considered.\cite{wedgewood1988model}.

\begin{figure}[htbp]
    \includegraphics[width=0.6\textwidth]{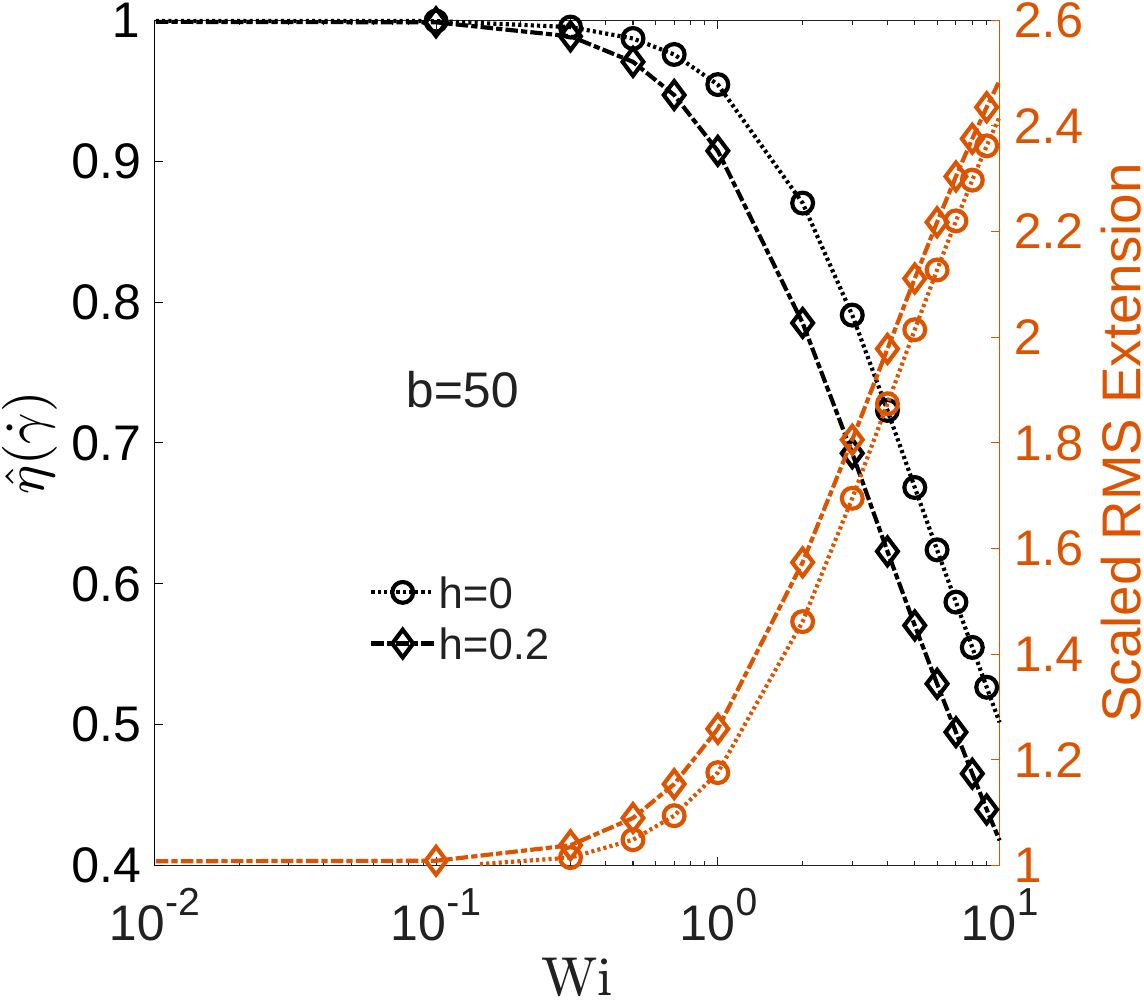}
\caption{Steady state value of the scaled-root mean square extension and normalized shear viscosity as function of Wi for two values of $h^\star = 0$ and 0.2 with extensibility parameter $b = 50$.}
\label{thinning_rms}
\end{figure}

The onset of shear thinning, on the other hand,  at small shear rates should be attributed to hydrodynamic interaction. 
To analyze this further,  shear viscosity is first normalized by its zero-shear value.    To calculate the zero-shear viscosity, defined as the shear viscosity in the limit ${\rm Wi} \to 0$, we extrapolated the shear viscosity obtained over the range ${\rm Wi} = 10^{-5}$ to $10^{-3}$. The extrapolated value at ${\rm Wi} = 10^{-8}$ was then taken as the zero-shear viscosity. The normalized shear viscosity, denoted with a hat  is given as:
\begin{equation}
\hat{\eta}(\dot{\gamma})=\frac{\eta(\dot{\gamma})-\eta_s}{\eta(0)-\eta_s}.\qquad
\end{equation}
In Fig. \ref{thinning_rms}, the normalized shear viscosity is plotted with respect to Wi  for $b=50$ and $h=0$ and $h=0.2$. 
For quantitative comparison, the onset of shear thinning is operationally defined as the Wi at which the normalized viscosity decreases to $95\%$ of its zero-shear value, i.e., $\hat{\eta}=0.95$. It can be observed from Fig. \ref{thinning_rms} that in the absence of hydrodynamic interactions ($h^\star=0$), the onset of shear thinning occurs close to ${\rm Wi}=1$. In contrast, for $h^\star=0.2$, this shifts to lower Wi (Wi$\approx$0.68), indicating an earlier onset of shear thinning. 
To examine the effect of hydrodynamic interactions (HI) on the microstructural dynamics of the dumbbell, we also plot the root mean square (RMS) extension, defined as ${\rm RMS\text{-}extension}=\sqrt{\langle \psi Q^2 \rangle/\langle \psi \rangle}$, where $\psi$ denotes the configurational probability density. 
The RMS extension is normalized by its equilibrium value, obtained by extrapolation using the same procedure as that employed for the zero-shear viscosity,
and plotted as a function of ${\rm Wi}$ in Fig. \ref{thinning_rms}.
Figure \ref{thinning_rms} shows that, at low ${\rm Wi}$, the RMS extension in the presence of hydrodynamic interactions is slightly larger than that in their absence. This can be attributed to the fact that, in the low-${\rm Wi}$ regime, the polymer remains predominantly in a coiled state and the imposed flow is not sufficiently strong to induce significant stretching of the chain.

However, hydrodynamic interactions introduce correlated motion between beads resulting in a slight expansion of the coil even under weak flow. On the other hand, at high Wi, the imposed flow dominates, and the polymer is stretched strongly, while hydrodynamic interactions become less effective. RMS extension increases sharply and approaches its finite extensibility limit, which can also be comprehended from the figure.
   \section{Conclusion\label{conclusion_hyd}}
A discrete deterministic kinetic method, based on lattice Boltzmann method, has been developed for homogeneous dilute polymer solution to incorporate the effect of hydrodynamic interactions. The scheme is validated against Brownian dynamics simulations and successfully captures the expected changes in material functions arising due to hydrodynamic interactions. 
In order to provide a   comparison with  BD,   we present  the computational time required 
to reach $t=10 \tau_{\rm R}$ with different resolutions for present method and different number of trajectories for BD in Table \ref{CPU-time}. The simulations are performed on the Intel(R) Core(TM) Ultra 7 155H 16 cores@4.8 GHz. The configuration  primarily used for comparison in this study is highlighted in bold in the Table \ref{CPU-time}. 
As can be seen from the Table \ref{CPU-time}, that this  method requires significantly little computational time compared to its stochastic counterpart (BD), which needs to compute the square-root of the diffusion tensor at every time
step.  
\begin{table}[h]
\begin{tabular}{|l|l||l|l|}
\hline
 \multicolumn{2}{|l|}{BD (dt=0.001) }&\multicolumn{2}{l|}{Present}\\
\cline{1-4}
 No. of trajectories &  CPU-time &No. of grid points &CPU-time\\
\hline\hline
 $10^5$&3 min 28 s&$20^3$&0 min 1.42 s\\
 $10^6$& 33 min 32 s& ${ 24^3}$& 0 min 3.23 s\\
 ${\bf 10^7}$& {\bf 5 hr 45 min 1 s}&${\bf 32^3}$&{\bf 0 min 10.8 s}\\
 $10^8$&  56 hr 39 min 53 s&$50^3$& 1 min 8.6 s\\
   \hline
\end{tabular}
\caption{\label{CPU-time} CPU time required to reach $t=10 \tau_{\rm R}$
at different resolution of present scheme as well as 
different number of trajectories for BD at fixed dt=0.001.
}
\end{table}
It is worth pointing out here that the Brownian dynamics code uses an explicit Euler time integration scheme, which can be   optimized  either by using the variance reduced scheme introduced by \"{O}ttinger \cite{ottinger1996stochastic} or by using the semi-implicit
scheme introduced by Ref.\cite{jendrejack2000hydrodynamic,jendrejack2002stochastic}. At the same time, the present scheme is also subjected to further optimization by using  proper vectorization schemes.
\begin{figure}
 \includegraphics[scale=0.5]{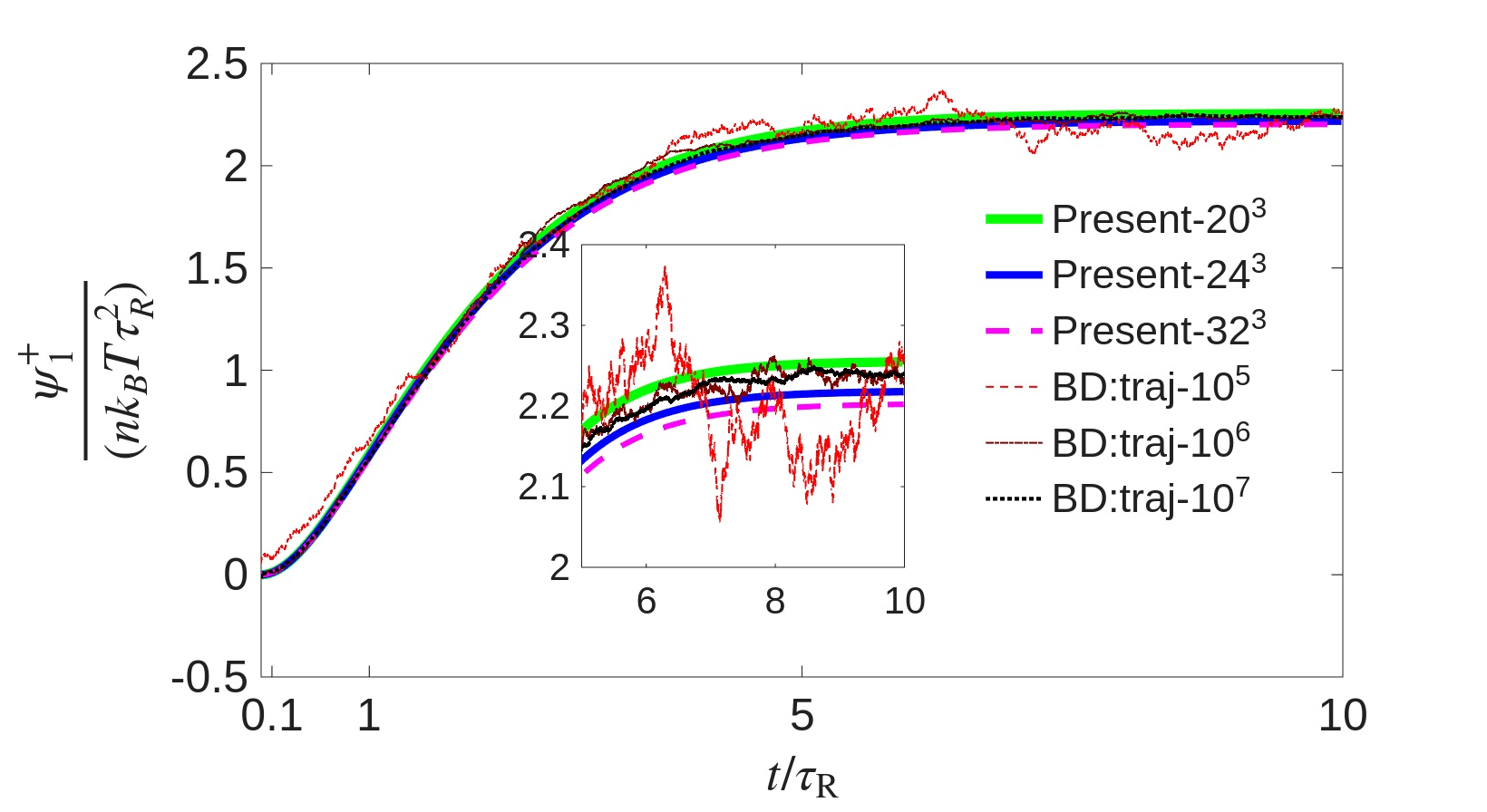}
 \label{comparison}
 \caption{Time evolution of first normal stress coefficient at Wi=0.3 and $h^\star = 0.15$ at different resolution of present scheme as well as different number of trajectories for BD.}
 \label{com_com}
 \end{figure}
Since in  the Brownian dynamics simulations, the statistical error is inversely proportional to the number of trajectories used,
the method becomes highly expensive in the limit ${\rm Wi}\to 0$.
Therefore, we wish to comment on the choice of number of trajectories used in BD simulations using Fig. \ref{com_com} where  the time dynamics of first normal stress at  Wi=0.3 is plotted.
It can be seen from the figure that at least $10^7$ trajectories are required to obtain the time evolution of the first normal stress coefficient with reasonably small statistical error. In contrast, the present scheme converges rapidly to the correct value with a comparatively small number of grid points ($\approx 20^3$). It should be noted that even with $10^7$ number of trajectories, BD
value for first normal stress coefficient contains some statistical fluctuation. Finally, we emphasize that the present methodology provides direct access to the distribution function, thereby enabling a detailed characterization of higher-order moments in the presence of hydrodynamic interactions. Such information can offer valuable physical insight and may aid in the development of more accurate and efficient constitutive models \cite{kaushal2025lpolymer}.
\section{Acknowledgment}
SA acknowledges the support via Abdul Kalam Technology Innovation National Fellowship (Grants No. INAE/SA/4784).
\appendix
\section{Chapman-Enskog Expansion}
In this section, Chapman-Enskog (CE) analysis is performed to recover the slow dynamics governed by the Smoluchowski equation for the configuration distribution from the full phase description in momentum space.  The Chapman–Enskog (CE) expansion is a perturbative method used to extract transport coefficients from a kinetic description. In this approach, only the conserved variable—here the configuration probability density $\psi$-is not expanded, while the non-conserved quantities, such as momentum ($\psi {\pmb u} $) and the stress tensor (${\pmb P}$), are expanded in powers of a small parameter $\epsilon $ as follows:
\begin{align}
\psi  u_\alpha&=\psi v_\alpha^{\rm HI}+\epsilon \psi  u_\alpha^{(1)}+\epsilon^2 \psi  u_\alpha^{(2)}+.....\label{u_correct}\\
 P_{\alpha\beta}&= \Theta_{\alpha\beta}+\epsilon P_{\alpha\beta}^{(1)}+\epsilon^2 P_{\alpha\beta}^{(2)}+.....
\end{align}
Furthermore,  the time derivative of any quantity $\phi$ is expanded as \cite{chapman1991mathematical}:
\begin{equation}
    \frac{\partial \phi}{\partial t}=\frac{\partial^{(0)} \phi}{\partial t}+\epsilon\frac{\partial^{(1)} \phi}{\partial t}+\epsilon^2\frac{\partial^{(2)} \phi}{\partial t}
\end{equation}
The multiscale expansion of the time derivative is substituted into the conservation equations (Eq.~\eqref{moment_eq}), which are obtained from the zeroth- and first-order moments of the Boltzmann–BGK equation (Eq.~\eqref{single_relax_grad}). The O(1) dynamics of $\psi$ and $\psi u_\alpha$ are then obtained by equating the coefficients of $\epsilon^0$, yielding:
\begin{align}
\frac{\partial^{(0)}\psi}{\partial t}+\frac{\partial v_\alpha^{\rm HI}}{\partial \dot{Q}_\alpha}&=0\label{psi0}\\
\frac{\partial^{(0)}\psi v_\alpha^{\rm HI}}{\partial t}+\frac{\partial \Theta_{\alpha\beta}}{\partial {Q}_\alpha}&=-\psi u_\alpha^{(1)}
\end{align}
Using the fact that  ${\pmb v}^{\rm HI}$ is an imposed velocity, together with the form of  ${\pmb \Theta}=\psi({\pmb \delta} -{\pmb\Omega}^{\rm RPY})$  and the divergence-free nature of ${\pmb\Omega}^{\rm RPY}$ , we obtain:
\begin{equation}
-\psi u_\alpha^{(1)}=({ \delta_{\alpha\beta}} -{\Omega^{\rm RPY}_{\alpha\beta}})\frac{\partial\psi}{\partial Q_\beta}
\end{equation}
Substituting the above expression into Eqs.~\eqref{u_correct} and \eqref{psi0}yields the following continuity equation:
\begin{equation}
\frac{\partial\psi}{\partial t}+\frac{\partial v_\alpha^{\rm HI}}{\partial \dot{Q}_\alpha}=\epsilon\frac{\partial}{\partial {Q}_\alpha}\left[({ \delta_{\alpha\beta}} -{\Omega^{\rm RPY}_{\alpha\beta}})\frac{\partial\psi}{\partial Q_\beta}\right]+O(\epsilon^2)
\end{equation}
which is the desired Smoluchowski equation. 
\section{Grad-distribution function}
Grad proposed expressing the distribution function $f({\pmb c},{\pmb x},t)$, which depends on particle velocity ${\pmb c}$, position ${\pmb x}$, at a given time $t$, as a series expansion in orthogonal Hermite polynomials. The number of Hermite coefficients retained in the expansion depends on the desired level of accuracy, with the complete (infinite) set of coefficients being equivalent to the distribution function $f$ itself \cite{grad1949kinetic}. The Grad’s distribution function ($f^{\rm Grad}$) is characterized by the values of relevant moments, including not only the locally conserved fields but also those that evolve more slowly than the others, such as the stress tensor. 
In most cases, the relevant variables are mass density, $\rho$, momentum density, $\rho {\pmb u}$ and the stress-tensor, ${\pmb P}$, defined as 
\begin{equation}
\int f^{\rm Grad} d{\pmb c}=\rho; \qquad \int f^{\rm Grad} c_\alpha \, d{\pmb c}=\rho  u_\alpha; \qquad \int f^{\rm Grad} c_\alpha c_\beta\, d{\pmb c}= P_{\alpha\beta},
\label{moment_grad}
\end{equation}
The expansion of distribution function in terms of Hermite polynomials is:
\begin{equation}
    f^{\rm Grad}=\frac{1}{(2\pi)^\frac{3}{2}}\exp{\left(\frac{-c^2}{2}\right)} \sum_{n=0}^{N}\frac{1}{n!} a^{(n)}{ H}^{(n)}({\bf c})
\end{equation}
where $H^{(n)}$ are the orthogonal Hermite tensors and $N$ defines the order of expansion. To obtain the momentum equation, it is necessary to take $N\geq 2$. Therefore, the first three required Hermite polynomials are:
\begin{eqnarray}
H^{(0)}({\bf c})=1; \qquad
H_\gamma^{(1)}({\bf c})=c_\gamma; \quad\textrm{and}\quad
H_{\gamma\theta}^{(2)}({\bf c})=c_{\gamma}c_{\theta}-\delta_{\alpha\theta} \label{hermite}
\end{eqnarray}
Therefore, the Grad-distribution  can be expressed as:
\begin{equation}
    f^{\rm Grad}=\frac{1}{(2\pi)^\frac{3}{2}}\exp{\left(\frac{-c^2}{2}\right)}\left[a^{(0)}+a^{(1)}_\gamma H_\gamma^{(1)}+\frac{1}{2}a^{(2)}_{\gamma\theta}H_{\gamma\theta}^{(2)} \right]
    \label{exp_grad}
\end{equation}
Using Eq. \eqref{moment_grad} and Eq. \eqref{exp_grad}, we get
\begin{equation}
a^{(0)}=\rho; \qquad 
a^{(1)}_\gamma=\rho u\gamma\quad\textrm{and}\quad
a^{(2)}_{\gamma\theta}=P_{\gamma\theta}-\rho\delta_{\gamma\theta}\label{a1_grad}
\end{equation}
It can be observed from Eqs. \eqref{hermite}, \eqref{exp_grad} and \eqref{a1_grad} that:
\begin{equation}
    {a}^{(n)}=\int f^{\rm Grad} H^{(n)} d{\pmb c}
\end{equation}
Therefore, final form of the distribution function can be written as:
\begin{equation}
    f^{\rm Grad}=w({\pmb c})\left[\rho+\rho u_\gamma c_\gamma+\frac{1}{2}(P_{\gamma\theta}-\rho\delta_{\gamma\theta})(c_{\gamma}c_{\theta}-\delta_{\alpha\theta}) \right]
    \label{final_grad}
\end{equation}
where $w({\pmb c})=\frac{1}{(2\pi)^\frac{3}{2}}\exp{\left(\frac{-c^2}{2}\right)}$.
The above equation represents the projection of the  distribution function onto to the finite subspace spanned by the lower order Hermite polynomial. 
To discretize this, one works with the the discrete velocities ${\pmb c}_i$ with $i = 1, ..., N$ taken as the nodes and $\omega_i$
are the associated weights of Gauss-Hermite quadrature. The Gaussian quadrature approximation of any integral of a function $\phi ({\pmb c})$ is
\begin{equation}
    \int w({\pmb c}) \phi ({\pmb c}) d{\pmb c}=\sum_{i=1}^{N}
\omega_i\phi ({\pmb c}_i).
\end{equation}
Since $f^{\rm Grad}$ is of the form $ w({\pmb c}) \phi({\pmb c})$, therefore the Hermite coefficients can be written as:
\begin{equation}
a_n=\sum_{}^{} \frac{\omega_i}{w({\pmb c_i})} f^{\rm Grad}({\pmb c}_i)H^{(n)}({\pmb c}_i)
\end{equation}
In LB literature for convenience, the discrete form of Maxwell-Boltzmann distribution function is defined as:
\begin{equation}
    f_i^{\rm Grad}=\frac{\omega_i f^{\rm Grad}({\pmb c}_i)}{w({\pmb c_i})}
\end{equation}
Therefore, the Hermite coefficients becomes
\begin{equation}
a_n=\sum_{}^{} f_i^{\rm Grad}H^{(n)}({\pmb c}_i)
\end{equation}
and the required moments become:
\begin{equation}
    \rho=\sum_i f_i^{\rm Grad};\qquad \rho { u}_\alpha=\sum_i {\pmb c_\alpha}_if_i^{\rm Grad}\quad\textrm{and}\quad  {P}_{\alpha\beta}=\sum_i {\pmb c_\alpha}_i {\pmb c_\beta}_if_i^{\rm Grad}
\end{equation}
Finally, discrete form of Grad distribution function becomes:
\begin{equation}
    f_i^{\rm Grad}=\omega_i\left[\rho+\rho u_\gamma c_{i\gamma}+\frac{1}{2}(P_{\gamma\theta}-\rho\delta_{\gamma\theta})(c_{i\gamma}c_{i\theta}-\delta_{\alpha\theta}) \right]
    \label{final_grad_discrete}
\end{equation}
\section*{References}
 \bibliographystyle{unsrt}
   \bibliography{refer}
\end{document}